\documentclass[namedreferences]{kluwer}

\usepackage[]{graphicx}

\newdisplay{guess}{Conjecture}

\newcommand{\araa}   {ARA\&A}
\newcommand{\aap}    {A\&A}
\newcommand{\apj}    {ApJ}
\newcommand{\apjl}   {ApJ}
\newcommand{\mnras}  {MNRAS}

\begin{document}
\begin{article}

\begin{opening}         

\title{Collimation of a Central Wind by a Disk-Associated Magnetic Field} 

\author{Sean \surname{Matt}\thanks{CITA National Fellow}
\email{matt@physics.mcmaster.ca}} 
\institute{Physics \& Astronomy Department; McMaster University}

\author{R. M. \surname{Winglee} \email{winglee@geophys.washington.edu}}
\institute{Earth \& Space Sciences; University of Washington}
\author{K.-H. \surname{B\"ohm} \email{bohm@astro.washington.edu}}
\institute{Astronomy Department; University of Washington}

\runningauthor{Matt, Winglee, \& B\"ohm}
\runningtitle{Collimation by a Disk Field}


\begin{abstract}

We present the results of time-dependent, numerical
magnetohydrodynamic simulations of a realistic young stellar object
outflow model with the addition of a disk-associated magnetic field.
The outflow produced by the magnetic star-disk interaction consists of
an episodic jet plus a wide-angle wind with an outflow speed
comparable to that of the jet (100--200 km s$^{-1}$).  An initially
vertical field of $\ll 0.1$ Gauss, embedded in the disk, has little
effect on the wind launching mechanism, but we show that it collimates
the entire flow (jet + wide wind) at large (several AU) distances.
The collimation does not depend on the polarity of the vertical field.
We also discuss the possible origin of the disk-associated field.

\end{abstract}


\end{opening}           

\section{Introduction}

Prevalent theoretical models for winds launched from accretion disks
(see, e.g., \opencite{koniglpudritz00} for a review) hold that the
final wind velocity is of the order of the Keplerian rotational
velocity of the launch point.  The high velocity of observed jets from
young stellar objects (YSO's) suggests that they originate from a deep
potential well \cite{kwantademaru88}, requiring the launching region
to be less than an AU in extent, for reasonable parameters.  We
therefore adopt the view that these jets are launched from a region
within several stellar radii.  The observations require that the flows
are launched initially with large opening angles and become collimated
within a few 10's of AU \cite{eisloffelea00}.  Since the winds become
collimated along the rotational axis of the accretion disk, the
collimation process must be associated with the disk.  In this work,
we explore one possible explanation, that disk-associated, poloidal
magnetic fields collimate a central, fast, wide-angle wind into an
optical jet.

\section{The Central Wind}


\begin{figure}
\centerline{\includegraphics[width=29pc]{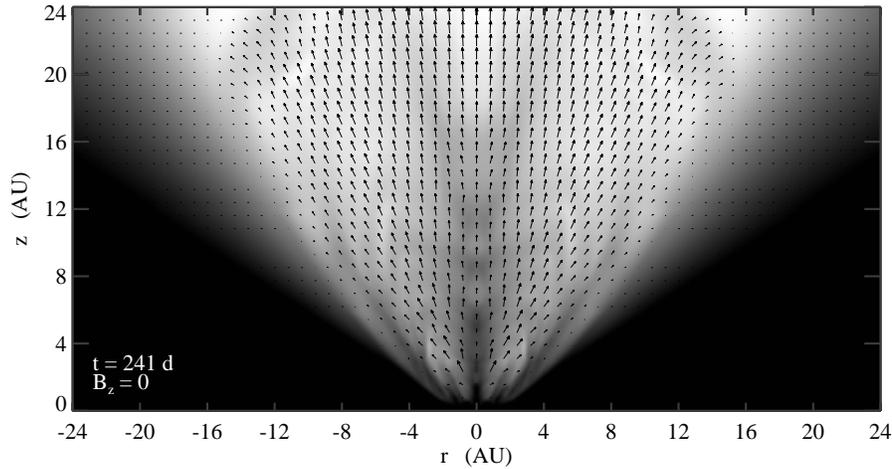}}

\caption[Jet and Wide Angle Flow]{Greyscale image of density 
(logarithmic) and velocity vectors from the case with no vertical
field at 241 days.  Log $\rho > -15.3$ gm cm$^{-3}$ is black, log
$\rho < -19.3$ gm cm$^{-3}$ is white.  From \inlinecite{matt3ea03}.
\label{fig_densbase}}

\end{figure}

\inlinecite{hayashi3ea96}, \inlinecite{goodson3ea97}, and
\inlinecite{millerstone97} studied a mechanism by which a rotating
star couples to the inner edge of a conductive accretion disk via a
stellar dipole magnetic field.  In this model, differential rotation
between the star and disk twists up the magnetic field, causing the
stellar magnetosphere to expand (or ``inflate'') above and below the
disk.  The field lines become effectively open, and material is
magnetocentrifugally launched \cite{blandfordpayne82} from the disk
inner edge and possibly from the star.  The disk inner edge spins down
and moves inward, forcing together the field lines from the star and
disk of opposite polarity, and instigating a reconnection of those
lines.  After reconnection, material at the disk inner edge accretes
via funnel flow along field lines onto the star.  The stellar
magnetosphere begins to expand outward and diffuse into the new inner
edge of the disk.  Again, differential rotation twists up the field,
and the process repeats. This process of episodic magnetospheric
inflation (EMI) regulates both the accretion and ejection of material
within the region of the disk inner edge.

Following the outflow to several AU, \inlinecite{goodson3ea99} showed
that the EMI mechanism produces an outflow consisting of a collimated,
highly structured jet and a wide angle wind.  The flow speed of both
components is $\gtrsim$ 100 km s$^{-1}$, in agreement with
observations \cite{reipurthbally01}.  Most of the kinetic energy and
mass is contained in the wide angle component of the wind.  Figure 1
contains the results of a magnetohydrodynamic (MHD) simulation and
illustrates the basic EMI outflow.

\inlinecite{mattea02} showed that a weak magnetic field, initially
aligned with the rotation axis and threading the accretion disk, did
not affect the launching of an outflow via the EMI mechanism.
However, they were unable to address the collimation of the outflow by
vertical fields at large distances because the size of their largest
simulation grid reached only to 0.75 AU.  Here, we follow up their
work and show that even a weak axis-aligned field will collimate the
entire EMI outflow.  For this work, we use the 2.5D MHD code of
\inlinecite{mattea02}, which solves the ideal MHD equations with the
added physics of gravity and Ohmic diffusion on a group of
cylindrical, nested grids.


\begin{figure}
\centerline{\includegraphics[width=29pc]{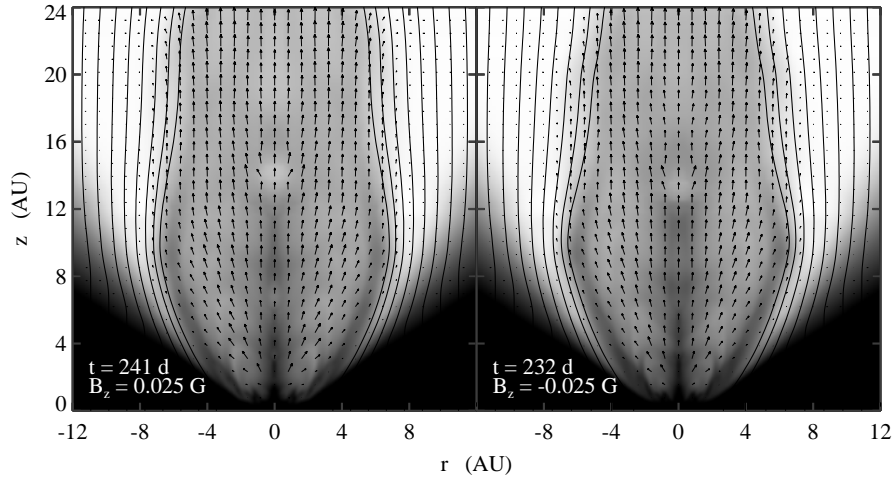}}

\caption[Wind Collimated by 0.025 Gauss Field]{Greyscale images of
density (logarithmic), magnetic field lines, and velocity vectors from
the case with a 0.025 Gauss vertical field.  The two panels represent
simulations that are identical except for the polarity of the vertical
field, and the time in days is shown in each panel.  The greyscale is
identical to Figure 1.  From \inlinecite{matt3ea03}.
\label{fig_densweak}}

\end{figure}

\section{Collimation by a Disk Field}

Figure 2 shows the results of two MHD simulations that include the
effect of an initially vertical magnetic field of 0.025 Gauss
threading the accretion disk.  The simulations in Figure 2 are
otherwise identical to that of figure 1.  The figure shows that the
wide angle component of the wind becomes collimated to a jet radius of
$\sim 8$ AU (where the wind kinetic energy density roughly equals the
vertical magnetic energy density).  This mechanism may be responsible
for producing more powerful and physically broader jets than by the
EMI mechanism alone.

Large-scale disk fields, if present, may be generated by disk currents
(as in \opencite{spruit3ea97}) and/or could be embedded in the surface
of the disk and carried outward in a disk wind
\cite{blandfordpayne82,kwantademaru88,ouyedpudritz97}, though a disk
wind is not yet included in our simulations.  In a realistic accretion
disk, it is possible that any field present in the disk will be
disordered.  A disk wind would therefore be threaded by a magnetic
field with polarity reversals at irregular intervals.  Such direction
reversals do not affect the qualitative behavior of the field
\cite{tsinganosbogovalov00}, so that, whether the vertical field has a
constant or chaotic polarity, it will always act to collimate the
flow.

\acknowledgements

This research was supported by NSF grant AST-9729096 and by NSERC,
McMaster University, and CITA through a CITA National Fellowship.



\end{article}
\end{document}